\def\figsize{9.5cm}

\def\rn{}
\def\nn#1 #2{#2. #1}				
\def\nnn#1 #2 #3{#2. #3. #1}			
\def\nnnn#1 #2 #3 #4{#2. #3. #4 #1}		
\def\nnnnn#1 #2 #3 #4 #5{#2. #3. #4 #5. #1}	
\def\dualand{ and\hbox{ }}				
\def\multiand{, and\hbox{ }}				
\def\rf#1;#2;#3;#4;#5 {{\frenchspacing\par\rn#1, #3 {\bf #4}, #5 (#2). \par}}
\def\rrf#1;#2;#3;#4;#5 {{\frenchspacing\rn#1, #3 {\bf #4}, #5 (#2);~}}
\def\rrrf#1;#2;#3;#4;#5 {{\frenchspacing\rn#1, #3 {\bf #4}, #5 (#2).}}
\def\rg#1;#2;#3;#4;#5;#6 {{\frenchspacing\par\rn#1, #3 {\bf #4}, #5 (#2). \par}}
\def\rfbook#1;#2;#3;#4;#5 {{\frenchspacing\par\rn#1, {\it #3} (#5, #4, #2).\par}}
\def\rfprep#1;#2;#3 {{\par\frenchspacing\rn#1, #3 (#2).\par}}
\def\rrfprep#1;#2;#3 {{\frenchspacing\rn#1, #3 (#2);~}}
\def\rrrfprep#1;#2;#3 {{\frenchspacing\rn#1, #3 (#2).}}
\def\rfproc#1;#2;#3;#4;#5;#6 {{\frenchspacing\par\rn#1 #2, in {\it #3}, ed. #4 (#5: #6)\par}}
\def\rfprocp#1;#2;#3;#4;#5;#6;#7 {{\frenchspacing\par\rn#1 #2, in {\it #3}, ed. #4 (#5: #6), p#7\par}}

\def\rg#1;#2;#3;#4;#5;#6 {\par\rn#1 #2, {\it #3}, {\bf #4}, #5 (``#6'') \par}
\def\rf#1;#2;#3;#4;#5 {\par\rn#1, {\it #3}, {\bf #4}, #5 (#2)\par}
\def\rfbook#1;#2;#3;#4;#5 {{\frenchspacing\par\rn#1, {\it #3} (#4: #5, #2)\par}}
\def\rfproc#1;#2;#3;#4;#5;#6 {{\frenchspacing\par\rn#1 #2, in {\it #3}, ed. #4 (#5: #6)\par}}
\def\rfprocp#1;#2;#3;#4;#5;#6;#7 {{\frenchspacing\par\rn#1 #2, in {\it #3}, ed. #4 (#5: #6), p#7\par}}
\def\rfprep#1;#2;#3  {{\par\rn#1, #3, #2\par}}
\def\rfprepp#1;#2;#3 {{\par\rn#1 #2, #3\par}}

\def\etal{{\frenchspacing\it et al.}}

\def\eg{{\frenchspacing\it e.g.}}

\def\bfk{\mbox{\bf k}}

\def\bfx{\mbox{\bf x}}

\newcommand{\be}{\begin{equation}}
\newcommand{\ee}{\end{equation}}
\newcommand{\ba}{\begin{eqnarray}}
\newcommand{\ea}{\end{eqnarray}}

\def\ga{\mathrel{\mathpalette\fun >}}
\def\fun#1#2{\lower3.6pt\vbox{\baselineskip0pt\lineskip.9pt
        \ialign{$\mathsurround=0pt#1\hfill##\hfil$\crcr#2\crcr\sim\crcr}}}
        
\def\beq#1{\begin{equation}\label{#1}}
\def\eeq{\end{equation}}
\def\beqa#1{\begin{eqnarray}\label{#1}}
\def\eeqa{\end{eqnarray}}

\def\spose#1{\hbox to 0pt{#1\hss}}
\def\simlt{\mathrel{\spose{\lower 3pt\hbox{$\mathchar"218$}}
     \raise 2.0pt\hbox{$\mathchar"13C$}}}
\def\simgt{\mathrel{\spose{\lower 3pt\hbox{$\mathchar"218$}}
     \raise 2.0pt\hbox{$\mathchar"13E$}}}
\def\simpropto{\mathrel{\spose{\lower 3pt\hbox{$\mathchar"218$}}
     \raise 2.0pt\hbox{$\propto$}}}

\def\ed{\end{document}}

\def\Om{\Omega_m}

\def\beq#1{\begin{equation}\label{#1}}
\def\eeq{\end{equation}}
\def\beqa#1{\begin{eqnarray}\label{#1}}
\def\eeqa{\end{eqnarray}}

\documentclass[twocolumn,amsmath,nofootinbib]{revtex4} 
\usepackage{url}
\begin{document}
\input{epsf.sty}
\input{psfig.sty}




\def\affilmrk#1{$^{#1}$}
\def\affilmk#1#2{$^{#1}$#2;}


\title{Observational approaches to understanding dark energy}

\author{Yun Wang}
\address{Homer L. Dodge Department of Physics \& Astronomy, Univ. of Oklahoma, 
440 W.~Brooks St., Norman, OK 73019, USA; wang@nhn.ou.edu}

\begin{abstract}

Illuminating the nature of dark energy is one of the most important
challenges in cosmology today. In this review I discuss
several promising observational approaches to understanding
dark energy, in the context of the recommendations by
the U.S. Dark Energy Task Force and the ESA-ESO Working Group
on Fundamental Cosmology.

\end{abstract}


  
\maketitle



\section{Introduction}

The discovery that the expansion of the universe
is accelerating today was first made in 1998 \cite{Riess98,Perl99}.
The evidence for the cosmic acceleration has strengthened over
time. Illuminating the nature of dark energy is one of the most exciting 
challenges in cosmology today.

The expansion history of the universe is described by the
Hubble parameter, $H(t)=(d\ln a/dt)=\dot a/a$, where $a(t)$ is the
cosmic scale factor, and $t$ is cosmic time.
The cosmological redshift, $z \equiv 1/a(t) -1$, is usually used
as the indicator for cosmic time, because it can be measured
for a given astrophysical object.
Fig.1 shows the Hubble parameter $H(z)$, as well as
$\dot a$, measured from current observational data \cite{WangPia07}.

\begin{figure} 
\vskip-1.3cm
\centerline{\epsfxsize=\figsize\epsffile{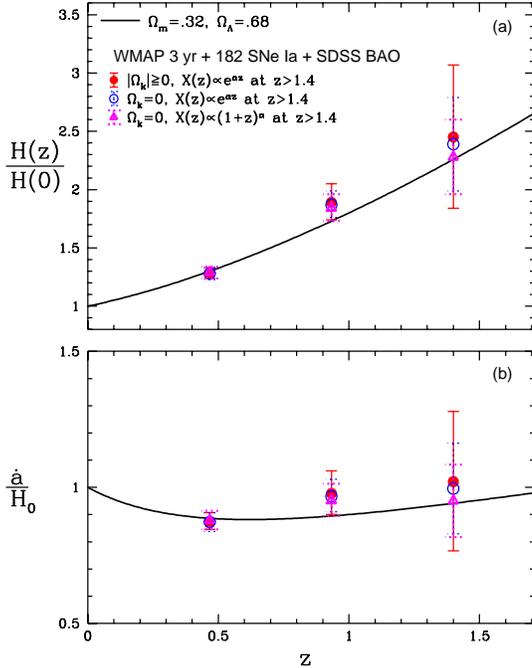}}
\vskip-0.3cm
\caption[1]{\label{Hz}\footnotesize%
Expansion history of the universe
measured from current data \cite{WangPia07}.
Data used:
Cosmic microwave background anisotropy (CMB) data from
WMAP 3 year observations \cite{Spergel06}; 
182 Type Ia supernovae (SNe Ia) (compiled by \cite{Riess07}, including 
data from the Hubble Space Telescope (HST) \cite{Riess07},
the Supernova Legacy Survey (SNLS) \cite{Astier06},
as well as nearby SNe Ia); 
Sloan Digital Sky Survey (SDSS) baryon acoustic 
oscillation measurement \cite{Eisen05}.
Note that $X(z)\equiv \rho_X(z)/\rho_X(0)$ in the figure legends.
}
\end{figure}

The observed cosmic acceleration could be due to an unknown 
energy component (dark energy, \eg, \cite{quintessence})), or a 
modification to general relativity (modified gravity, \eg, 
\cite{modifiedgravity,DGPmodel}). 
Ref.\cite{reviews} contains reviews with more complete 
lists of references of theoretical models.

The simplest explanation for the observed cosmic acceleration is
that dark energy is a cosmological constant (although it is
many orders smaller than expected based on known 
physics), and that gravity is not modified.
Fig.1 and Fig.2 show that a cosmological constant is
consistent with current observational data, although uncertainties 
are large.
For complementary approaches to analyzing current data, see, e.g., 
\cite{WangTegmark05,data_cur}.

\begin{figure} 
\psfig{file=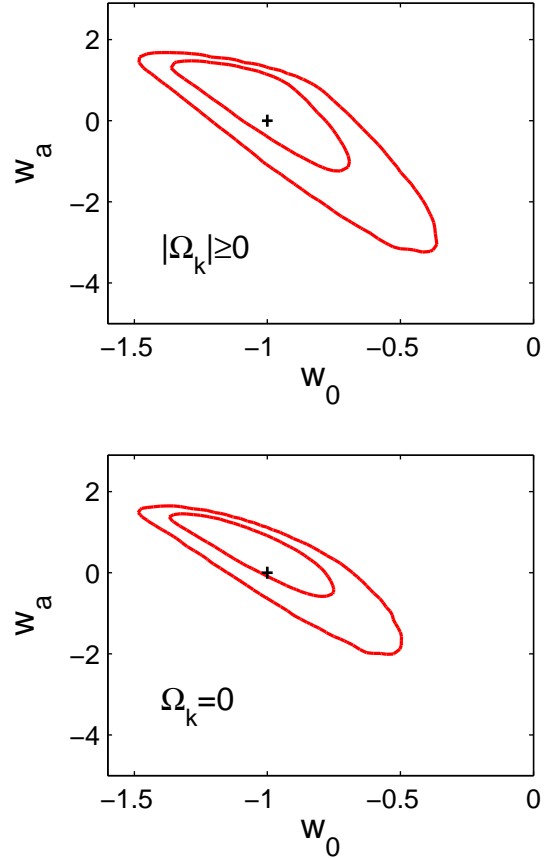,height=5in}
\vspace{-0.1in}
\caption{\footnotesize{Constraints on the dark energy equation of state 
$w_X(a)=w_0+w_a(1-a)$ \cite{Chev01}
using the same data as in
Fig.\ref{Hz}. 
A cosmological constant corresponds to $w_X(a)=-1$
(indicated by the cross in the figures). \cite{WangPia07}
}}
\label{fig:w0wp_now}
\end{figure}

I will discuss the general guidelines for dark energy search
in Sec.2, several promising observational methods in
Sec.3, and conclude with a summary
of current status and future prospects 
of dark energy experiments in Sec.4.

\section{General guidelines for dark energy search}

There are two fundamental questions that need to be answered
by dark energy search:\\
\noindent
(1) Is dark energy density constant in cosmic time?\\
\noindent
(2) Is gravity modified?\\
These questions can be answered by the precise and
accurate measurement of the 
dark energy density $\rho_X(z)$ as a function
of cosmic time [or the expansion history of
the universe, $H(z)$, see Eq.(\ref{eq:E(z)})], 
and the growth history of cosmic large scale structure, 
$f_g(z)$ [see Eqs.(\ref{eq:fg}) and (\ref{eq:fgdef})], 
from observational data.

Dark energy is often parameterized by
a linear equation of state $w_X(a)=w_0+w_a (1-a)$ \cite{Chev01}.
Because of our ignorance of the nature of dark energy,
it is important to make model-independent constraints
by measuring the dark energy density $\rho_X(z)$
[or the expansion history $H(z)$] as a free function
of cosmic time.
Measuring $\rho_X(z)$ has advantages over measuring dark energy
equation of state $w_X(z)$ as a free function; $\rho_X(z)$ is more
closely related to observables, hence is more tightly 
constrained for the same number of redshift bins 
used \cite{WangGarna,Tegmark02,WangFreese}.
Note that $\rho_X(z)$ is related to $w_X(z)$ as follows \cite{WangGarna}:
\begin{equation}
\label{eq:rhoprimew}
\frac{\rho_X(z)}{\rho_X(0)} = \exp\left\{ \int_0^z {\rm d}z'\, \frac{3
    [1+w_X(z')]}{1+z'} \right\}, 
\end{equation} 
{\it Hence parametrizing dark energy with $w_X(z)$ 
implicitly assumes that $\rho_X(z)$ does not
change sign in cosmic time.} 
This precludes whole classes of
dark energy models in which $\rho_X(z)$ becomes negative in the future
(``Big Crunch'' models, see \cite{Linde} for an example)\cite{WangTegmark04}.

If the present cosmic acceleration is caused by dark energy,
\be
E(z) \equiv \frac{H(z)}{H_0}=
\left[\Omega_m (1+z)^3 + \Omega_k (1+z)^2 +\Omega_X X(z)
\right]^{1/2},
\label{eq:E(z)}
\ee
where $X(z)\equiv \rho_X(z)/\rho_X(0)$. 
$H_0=H(z=0)$ is the Hubble constant.
$\Omega_m$ and $\Omega_X$ are the ratios of the matter and
dark energy density to the critical density $\rho_c^0=3H_0^2/(8\pi G)$,
and $\Omega_k=-k/H_0^2$ with $k$ denoting the curvature constant.
Consistency of Eq.(\ref{eq:E(z)}) at $z=0$ requires that 
$\Omega_m+\Omega_k+\Omega_X=1$.
Once $E(z)$ is specified, the evolution of matter density perturbations
on large scales, $\delta^{(1)}(\bfx,t)=D_1(t) \delta(\bfx)$ is determined
by solving the following equation for $D_1=\delta^{(1)}(\bfx,t)/\delta(\bfx)$, 
\be
\label{eq:fg}
D_1''(\tau) + 2E(z)D_1'(\tau) - {3\over 2}\Om (1+z)^3D_1 = 0,
\ee
where primes denote $d/d(H_0 t)$.
The linear growth rate 
\be
\label{eq:fgdef}
f_g(z) \equiv d\ln D_1/d\ln a.
\ee

Note that we have assumed that dark energy and dark matter are separate,
which is true for the vast majority of dark energy models
that have been studied in the literature. 
If dark energy and dark matter are coupled (a more complicated
possibility), or if dark energy and dark matter are unified
(unified dark matter models), Eq.(\ref{eq:fg}) would need 
to be modified accordingly.
Ref.\cite{Sandvik04} found the first strong evidence for the 
separation of dark energy and dark matter by ruling out a broad 
class of so-called unified dark matter models. They showed 
that these models produce oscillations or exponential blowup 
of the dark matter power spectrum inconsistent with observation.

In the simplest alternatives to dark energy,
the present cosmic acceleration is caused by a modification
to general relativity. 
Ref.\cite{modified_gravity_details} contains examples of
studies of observational signatures of modified gravity models.
The only rigorously worked example
is the DGP gravity model \cite{DGPmodel,DGP},
which can be described by a modified 
Friedmann equation\footnote{The validity of
the DGP model has been studied by \cite{DGP2}.}:
\be
H^2 - \frac{H}{r_0}=\frac{8\pi G \rho_m}{3},
\label{eq:DGP}
\ee
where $\rho_m(z)=\rho_m(0)(1+z)^3$. Soving the above equation gives 
\be
E(z)=\frac{1}{2} \left\{ \frac{1}{H_0 r_0}+
\sqrt{ \frac{1}{(H_0 r_0)^2} + 4 \Omega_m^0 (1+z)^3} \right\},
\label{eq:H(z)_DGP}
\ee
with $\Omega_m^0 \equiv \rho_m(0)/\rho_c^0$, $\rho_c^0 \equiv
3H_0^2/(8\pi G)$. The added superscript ``0'' in
$\Omega_m^0$ denotes that this is the matter fraction today
in the DGP gravity model.
Note that consistency at $z=0$, $H(0)=H_0$ 
requires that $H_0 r_0 = 1/(1-\Omega_m^0)$, so the
DGP gravity model is parametrized by a single parameter,
$\Omega_m^0$.
The linear growth factor in the DGP gravity model is
given by \cite{DGP}
\be
\label{eq:fg_DGP}
D_1''(\tau) + 2E(z)D_1'(\tau) - {3\over 2}\Om (1+z)^3D_1 
\left(1+ \frac{1}{3\alpha_{DGP}} \right)= 0,
\ee
where $\alpha_{DGP}=[1-2H_0r_0+ 2(H_0 r_0)^2]/(1-2H_0r_0)$.

The dark energy model equivalent of the DGP gravity model
is specified by requiring $8\pi G\rho_{de}^{eff}/3=H/r_0$.
Eq.(\ref{eq:DGP}) and the conservation of energy and
momentum equation, $\dot{\rho}_{de}^{eff}+3(\rho_{de}^{eff}
+p_{de}^{eff})H=0$, implies that
$w_{de}^{eff}=-1/[1+\Omega_m(a)]$ \cite{DGP}, where
$\Omega_m(a)  \equiv 8\pi G \rho_m(z)/(3 H^2)
= \Omega_m^0 (1+z)^3/E^2(z)$.
As $a\rightarrow 0$, $\Omega_m(a) \rightarrow 1$,
and $w_{de}^{eff} \rightarrow -0.5$.
As $a \rightarrow 1$, $\Omega_m(a) \rightarrow \Omega_m^0$,
and $w_{de}^{eff} \rightarrow -1/(1+\Omega_m^0)$.
This means that the matter transfer function for
the dark energy model equivalent of viable DGP gravity model
($\Omega_m^0<0.3$ and $w\leq -0.5$) are very close
to that of the $\Lambda$CDM model 
at $k\ga 0.001\,h\,$Mpc$^{-1}$.\cite{Ma99}

It is very easy and straightforward to integrate 
Eqs.(\ref{eq:fg}) and (\ref{eq:fg_DGP}) to obtain
$D_1(a)$ for dark energy models and DGP gravity models,
with the initial condition that for $a\rightarrow 0$, $D_1(a) = a$
(which assumes that the dark energy or modified gravity
are negligible at sufficiently early times).

The measurement of $H(z)$ or $\rho_X(z)$ allows us to determine
whether dark energy is a cosmological constant.
The measurement of $f_g(z)$ allows us to determine
whether gravity is mofidied.

\section{Observational methods}

I will discuss the use of Type Ia supernovae,
galaxy redshift surveys, weak lensing,
and galaxy clusters in probing dark energy.
These are recognized by the community as the most
promising methods for dark energy search.
Cosmic microwave background anisotropy (CMB) data 
and independent measurements of $H_0$ are
required to break the degeneracy between dark energy
and cosmological parameters (see e.g. \cite{WangPia07}), 
hence are important as well in dark energy search.

\subsection{Type Ia supernovae as dark energy probe}

The use of Type Ia supernovae (SNe Ia) 
is the best established method for probing dark energy,
since this is the method through which cosmic acceleration
has been discovered \cite{Riess98,Perl99}. The unique advantage of this
method is that it is independent of the clustering of
matter, and can provide a robust measurement of $H(z)$
through the measured luminosity distance as a function of redshift,
$d_L(z)=(1+z)\, r(z)$, where the comoving distance $r(z)$ 
from the observer to redshift $z$ is given by
\ba
\label{eq:rz}
&&r(z)=cH_0^{-1}\, |\Omega_k|^{-1/2} {\rm sinn}[|\Omega_k|^{1/2}\, \Gamma(z)],\\
&&\Gamma(z)=\int_0^z\frac{dz'}{E(z')}, \hskip 1cm E(z)=H(z)/H_0 \nonumber
\ea
where ${\rm sinn}(x)=\sin(x)$, $x$, $\sinh(x)$ for 
$\Omega_k<0$, $\Omega_k=0$, and $\Omega_k>0$ respectively.

\begin{figure} 
\vskip-1.3cm
\centerline{\epsfxsize=\figsize\epsffile{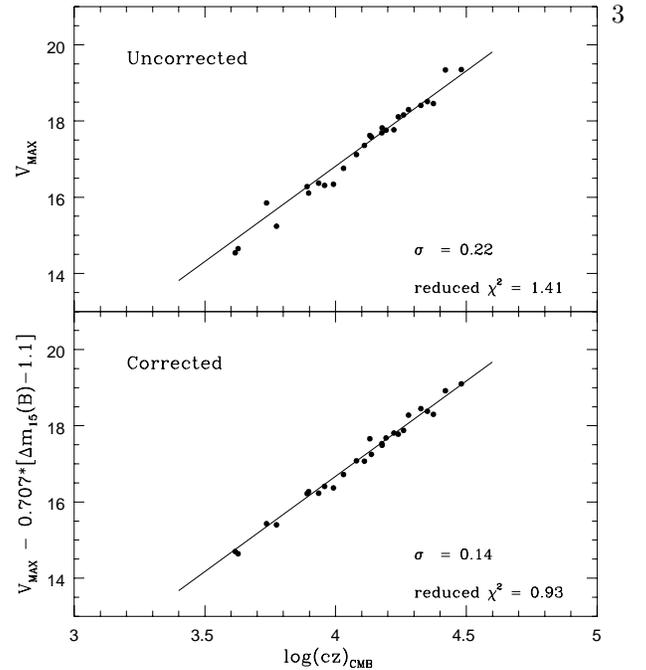}}
\vskip-0.3cm
\caption[1]{\label{fig:sn_cali}\footnotesize%
Hubble diagrams showing 26 SNe Ia
with $B_{max}-V_{max} \leq 0.20$ from the Calan/Tololo 
sample \cite{Hamuy96}.
This sample provided half of the data for the discovery of the 
cosmic acceleration in 1998 \cite{Riess98}. 
The solid lines indicate Hubble's law;
perfect standard candles (with $\sigma=0$) fall on these lines.
}
\end{figure}

A SN~Ia is a thermonuclear explosion that
completely destroys a carbon/oxygen white dwarf at
the Chandrasekher limit of 1.4 $M_{\odot}$.\footnote{The exception recently found by 
\cite{Howell06} is super-luminous
and can be easily separated from the normal SNe~Ia used for cosmology.} 
This is the reason SNe~Ia are so uniform in peak luminosity.
The first challenge to overcome when using SNe~Ia as cosmological
standard candles is properly incorporating the intrinsic scatter in SN~Ia peak luminosity.
The usual calibration of SNe~Ia reduces the intrinsic scatter in SN~Ia peak 
luminosity (Hubble diagram dispersion) to about 0.16~mag 
\cite{Phillips93,Riess95}.
The calibration techniques used so far are based on one observable parameter,
the lightcurve width, which can be parametrized either as $\Delta m_{15}$
(decline in magnitudes for a SN~Ia in the first 15 days after $B$-band maximum,
see \cite{Phillips93}), or a stretch factor (which linearly scales the time axis, 
see \cite{Goldhaber01}).
The lightcurve width is associated with the amount of $^{56}$Ni
produced in the SN~Ia explosion, which in turn depends on
when the carbon burning makes the transition from turbulent 
deflagration to a supersonic detonation \cite{Wheeler03}.
There may be additional physical parameters associated with
SN Ia lightcurves \cite{WangHall07} or spectra \cite{Branch}
that can further improve the calibration of SNe Ia.
Fig.\ref{fig:sn_cali} shows the homogeneity of SNe Ia \cite{Hamuy96}.
Fig.{\ref{fig:sn_07} shows the most recent homogeneous 
sample of SNe Ia \cite{Riess07}.
\begin{figure} 
\centerline{\epsfxsize=\figsize\epsffile{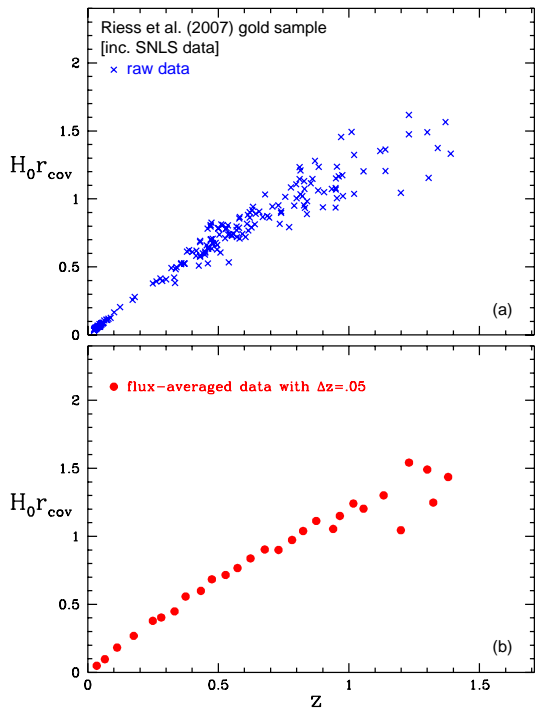}}
\vskip-0.3cm
\caption[1]{\label{fig:sn_07}\footnotesize%
The distance-redshift diagram of
the most recent homogeneous 
sample of SNe Ia compiled by \cite{Riess07},
including data from HST \cite{Riess07}, SNLS \cite{Astier06},
and nearby SNe Ia. This is an updated
version of Fig.1 in \cite{WangTegmark05}.
Here $r_{cov}=r(z)$ is the comoving distance of
an object at redshift $z$.
}
\end{figure}

The key to the efficient use of SNe Ia for probing dark nergy is
to obtain the largest possible unbiased sample of SNe Ia at the greatest
distances from the observer \cite{WangLovelace2001}.
This is achieved by an ultra deep survey of the same areas in
the sky every few days over at least one year \cite{Wang2000a}.
Fig.{\ref{fig:sn_deep}} compared an ultra deep supernova survey
\cite{Wang2000a} with a much shallower survey. Clearly,
a sufficiently deep supernova survey is required
to reconstruct the dark energy density $\rho_X(z)$ as
a free function of cosmic time (i.e., to measure $H(z)$ precisely).
\begin{figure} 
\centerline{\epsfxsize=\figsize\epsffile{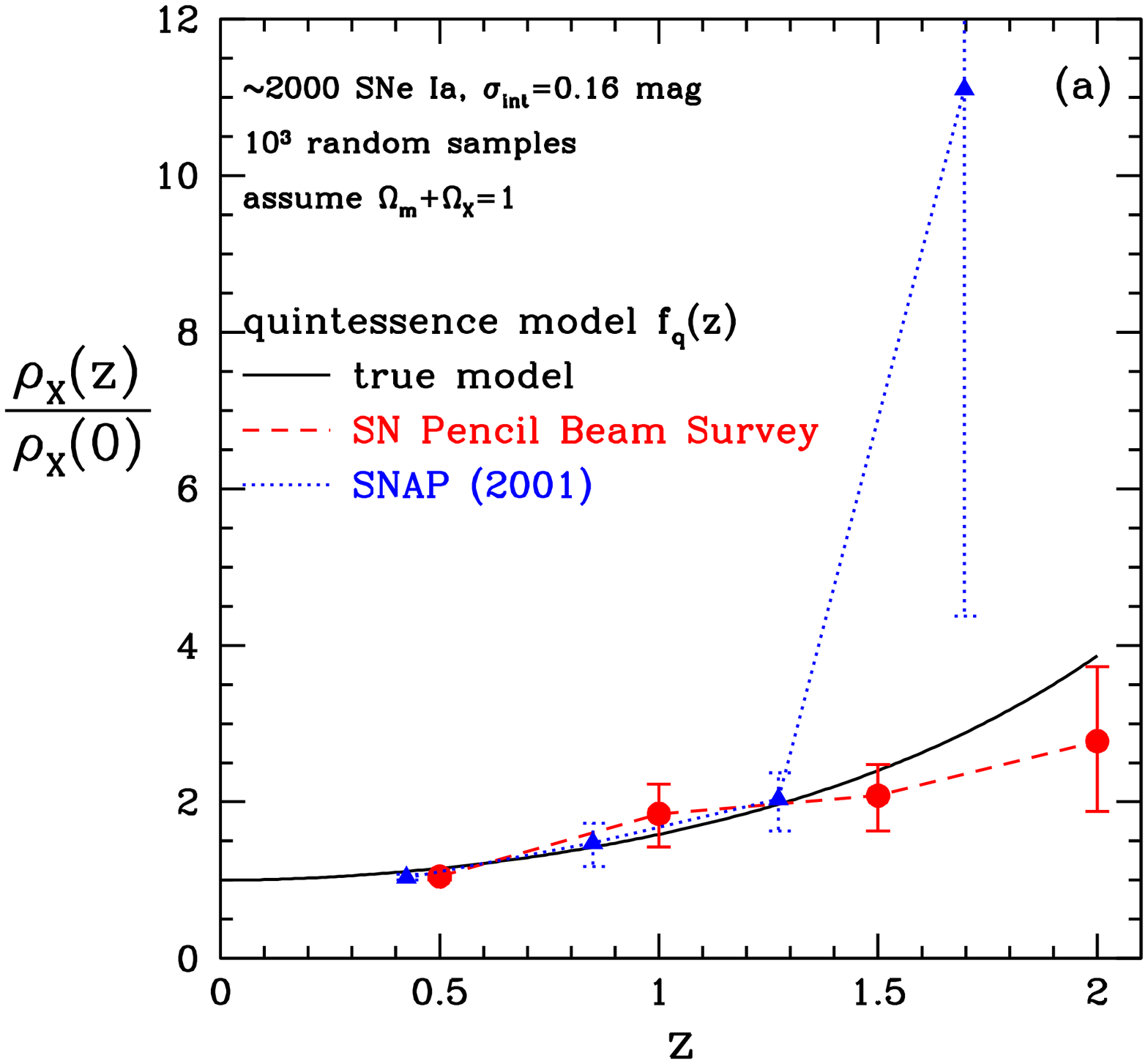}}
\centerline{\epsfxsize=\figsize\epsffile{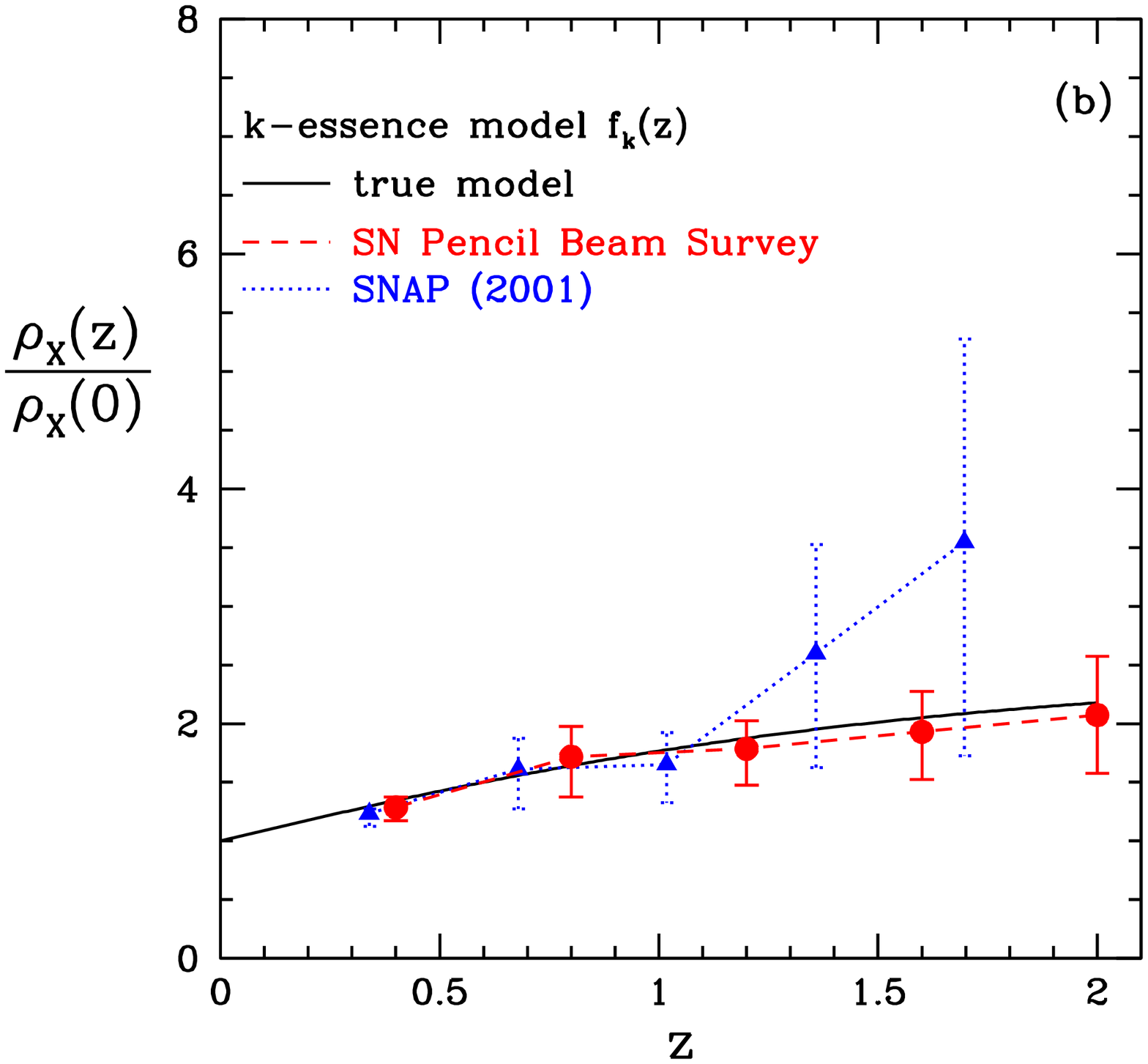}}
\vskip-0.3cm
\caption[1]{\label{fig:sn_deep}\footnotesize%
The comparison of an ultra deep supernova survey
\cite{Wang2000a} with a much shallower survey
in the reconstruction of the dark energy density $\rho_X(z)$ as
a free function of cosmic time \cite{WangLovelace2001}.
}
\end{figure}

The main systematic effects for using SNe Ia to probe dark energy
are: extinction by normal \cite{Card89}
or gray dust \cite{Aguirre99}, weak lensing amplification by cosmic
large scale structure \cite{sn_weak lensing}, and possible evolution in the 
peak luminosity of SNe Ia.

Gray dust, consisting of large dust grains,
is difficult to detect by its reddening and could mimic
the effect of dark energy \cite{Aguirre99}. 
Gray dust can be constrained quantitatively by the Cosmic Far 
Infrared Background \cite{Aguirre00},
with no evidence found in favor of gray dust so far.
Supernova flux correlation measurements can be used in 
combination with other lensing data to infer the level of dust 
extinction, and provide a viable method to 
eliminate possible gray dust contamination
in SN Ia data \cite{Zhang07}.

The extinction by normal dust can be corrected using
multi-band imaging data.
Recent data show that the apparent dust extinction of
SNe Ia is very different from the typical extinction law
due to Milky Way dust, possibly due to
the mixing of intrinsic SN Ia color variation with
dust extinction, or variations
in the properties of dust \citep{Conley07}.
The key to minimizing the systematic effect due to dust extinction
is to observe SNe Ia in the near infrared (NIR), since
dust extinction decreases with wavelength.

NIR observations of SNe Ia have an important
added advantage: SNe Ia are even better standard 
candles at NIR wavelengths \cite{KK04,Phillips06,WV07}.
Fig.{\ref{fig:sn_NIR}} shows the Hubble diagram of
SNe Ia in the NIR, {\it without} the usual lightcurve
width correction.
\begin{figure} 
\psfig{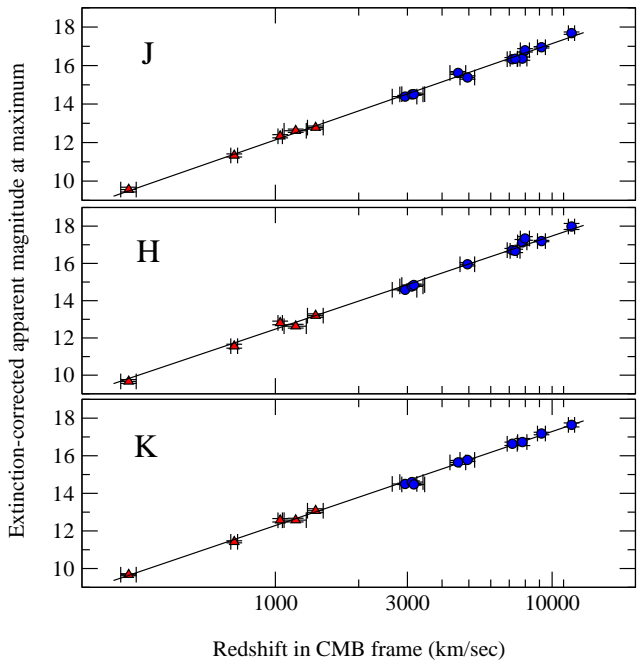}
\vspace{-0.1in}
\caption{\footnotesize{\footnotesize%
Hubble diagrams of SNe Ia in the NIR bands.
Note that these SNe Ia have only been corrected for
dust extinction; {\it no} corrections have been made
for lightcurve width. \cite{KK04}}}
\label{fig:sn_NIR}
\end{figure}

The weak lensing amplification of SNe Ia by cosmic
large scale structure can be modeled by a universal 
probability distribution function for 
weak-lensing amplification based on the measured
matter power spectrum \cite{Wang02}.
The effect of weak lensing on the SN Ia data
can be minimized through flux-averaging \cite{flux-averaging}.

The evolution in SN Ia peak luminosity could arise
due to progenitor population drift, since the most distant
SNe Ia come from a stellar environment very different 
(a much younger universe) than that of the nearby SNe Ia.
However, with sufficient statistics, we can subtype SNe Ia
and compare SNe Ia at high redshift and low redshift 
that are similar in both lightcurves and spectra,
thus overcoming the possible systematic effect due to
progenitor population drift \cite{Branch01}.

\subsection{Galaxy redshift survey as dark energy probe}

A magnitude-limited galaxy redshift survey can
allow us to measure the cosmic expansion history
$H(z)$ through baryon acoustic oscillations (BAO)
in the galaxy distribution, and the growth history of
cosmic large scale structure $f_g(z)$ through
independent measurements of redshift-space distortions and
the bias factor between the distribution of galaxies
and that of matter.\cite{Wang07}

The use of BAO as a
cosmological standard ruler is a relatively new
method for probing dark energy \cite{BG03,SE03,BAO},
but it has already yielded impressive constraints
on dark energy \cite{Eisen05}.

At the last scattering of CMB photons, the acoustic oscillations 
in the photon-baryon fluid became frozen, and imprinted their signatures
on both the CMB (the acoustic peaks in the CMB angular
power spectrum) and the matter distribution (the baryon acoustic
oscillations in the galaxy power spectrum).
Because baryons comprise only a small fraction of matter,
and the matter power spectrum has evolved significantly
since last scattering of photons, BAO are much smaller in amplitude
than the CMB acoustic peaks, and are washed out on small
scales.

BAO in the observed galaxy power spectrum have the characteristic 
scale determined by the comoving sound horizon at recombination, 
which is precisely measured by the CMB anisotropy data \cite{Spergel06}.
Comparing the observed BAO scales with the expected values
gives $H(z)$ in the radial direction, and $D_A(z)=r(z)/(1+z)$ (the angular
diameter distance) in the transverse direction.
Fig.{\ref{fig:BAO}} shows the BAO scale measurement
by the SDSS team, which is usually quoted in the form
of $A_{BAO}=0.469\,(n_S/0.98)^{-0.35} \pm 0.017$
(with $n_S$ denoting the power-law index of the
primordial matter power spectrum), to
be compared with the theoretical prediction of 
\be
\label{eq:A}
A = \left[ r^2(z_{BAO})\, \frac{cz_{BAO}}{H(z_{BAO})} \right]^{1/3} \, 
\frac{\left(\Omega_m H_0^2\right)^{1/2}} {cz_{BAO} },
\ee
with $z_{BAO}=0.35$. \cite{Eisen05}
\begin{figure} 
\centerline{\epsfxsize=\figsize\epsffile{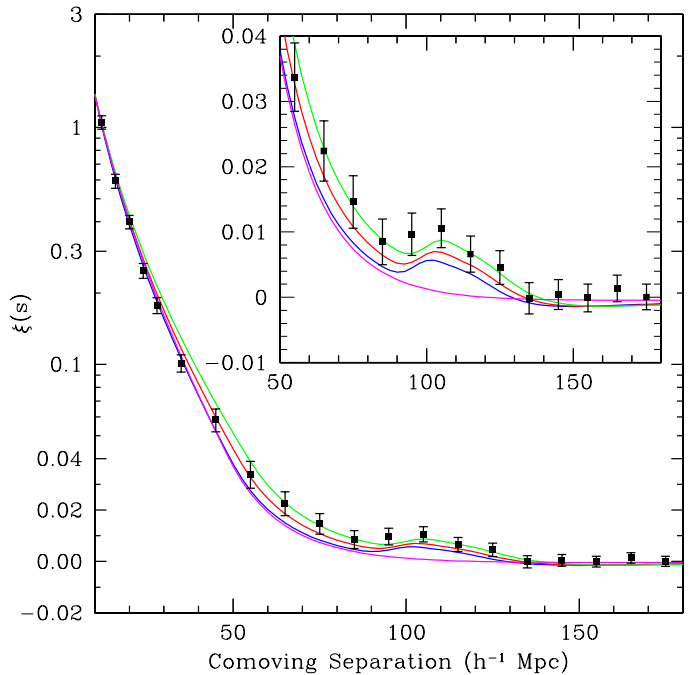}}
\vskip-0.3cm
\caption[1]{\label{fig:BAO}\footnotesize%
The galaxy correlation function measured from the SDSS data,
clearly showing a peak corresponding to the
BAO scale at $\sim$ 100$\,h^{-1}$Mpc \cite{Eisen05}.
}
\end{figure}

A magnitude-limited galaxy redshift survey can
allow us to measure both $H(z)$ and $f_g(z)$ \cite{Guzzo07,Wang07}.
The measurement of $f_g(z)$ can be obtained through
independent measurements of redshift-space distortion parameter
$\beta=f_g(z)/b$ \cite{Kaiser87} and the bias
parameter $b(z)$ (which describes how light traces mass) \cite{Guzzo07}.
The parameter $\beta$ can be measured directly from galaxy redshift 
survey data by studying the observed redshift-space correlation function
\cite{2dFbeta,beta}. 
We can assume that the galaxy density perturbation $\delta_g$
is related to the matter density perturbation $\delta(\bfx)$
as follows \cite{Fry93}:
\be
\delta_g= b \delta(\bfx)+ b_2 \delta^2(\bfx)/2.
\ee
The galaxy bispectrum is
\ba
\langle \delta_{g\bfk_1} \delta_{g\bfk_2} \delta_{g\bfk_1}\rangle
&=& (2\pi)^3 \left\{P_g(\bfk_1) P_g(\bfk_2)\left[J(\bfk_1,\bfk_2)/b
+b_2/b^2\right] \right.\nonumber\\
& & \left. +cyc.\right\} \delta^D(\bfk_1+\bfk_2+\bfk_3),
\ea
where $J$ is a function that depends on the shape of the
triangle formed by ($\bfk_1$, $\bfk_2$, $\bfk_3$) in
$\bfk$ space, but only depends very weakly on cosmology \cite{bias}.
Ref.\cite{bias} developed the method for measuring $b(z)$
from the galaxy bispectrum, which was applied by \cite{2dFbias}
to the 2dF data.
Independent measurements of $\beta(z)$ and $b(z)$
have only been published for the 2dF data 
\cite{2dFbeta,2dFbias}.

Fig.{\ref{fig:Hzfg}} shows how well a magnitude-limited
NIR galaxy redshift survey
covering $>$10,000 square degrees and $0.5<z<2$ can constrain
$H(z)$ and $f_g(z)$, compared with current data \cite{Wang07}.
Fig.{\ref{fig:Hzfg}}(b) shows the $f_g(z)$ for a modified
gravity model (the DGP gravity model)
with $\Omega_m^0=0.25$ (solid line), as well as a dark energy model
that gives the same $H(z)$ for the same $\Omega_m^0$ (dashed line).
The cosmological constant model from Fig.{\ref{fig:Hzfg}}(a) 
is also shown (dotted line).
Clearly, current data can not differentiate between dark energy
and modified gravity.
A very wide and deep galaxy redshift survey provides
measurement of $f_g(z)$ accurate to a few percent
[see Fig.{\ref{fig:Hzfg}}(b)]; this will allow an unambiguous distinction
between dark energy models and modified gravity models
that give identical $H(z)$ [see the solid and dashed lines
in Fig.{\ref{fig:Hzfg}}(b)].
For a linear cutoff given by $\sigma^2(R)= 0.35$ (or 0.2),
a survey covering 11,931$\,$(deg)$^2$ would rule out the
DGP gravity model that gives the same
$H(z)$ and $\Omega_m^0$ at 99.99\% (or 95\%) C.L.;
a survey covering 13,912$\,$(deg)$^2$ would rule out the
DGP gravity model at 99.999\% (or 99\%) C.L. \cite{Wang07}.
\begin{figure} 
\centerline{\epsfxsize=\figsize\epsffile{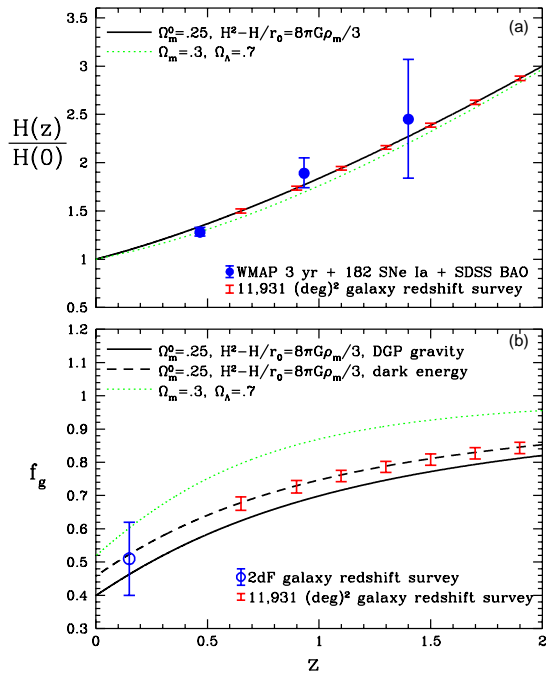}}
\vskip-0.3cm
\caption[1]{\label{fig:Hzfg}\footnotesize%
Current and expected future measurements of the 
cosmic expansion history $H(z)=H_0 E(z)$ and the growth rate of 
cosmic large scale structure 
$f_g(z)=d\ln \delta/d\ln a$ ($\delta=(\rho_m-\overline{\rho_m})/
\overline{\rho_m})$, $a$ is the cosmic scale factor).
The future data correspond to a magnitude-limited
NIR galaxy redshift survey
covering $>$10,000 square degrees and $0.5<z<2$. 
If the $H(z)$ data are fit by both a DGP gravity model and
an equivalent dark energy model that predict the same
expansion history, a survey area of 11,931$\,$(deg)$^2$ is required
to rule out the DGP gravity model at the 99.99\% confidence level.
\cite{Wang07}
}
\end{figure}

The systematic effects of BAO as a standard ruler are: bias between
luminous matter and matter distributions, nonlinear effects, and 
redshift distortions \citep{BG03,SE03}. Cosmological N-body
simulations are required to quantify these effects
\citep{Angulo05,SE05,Springel05,White05}.
Ref.\cite{nonlinear1} shows that nonlinear effects can be accurately
taken into account. Ref.\cite{nonlinear2} shows that
the BAO signal is {\it boosted} when bias, nonlinear effects,
and redshift distortions are properly
included in the Hubble Volume simulation.

\subsection{Weak lensing as dark energy probe}

Weak lensing usually refers to the statistically correlated 
image distortions of background galaxies due to the foreground
matter distribution, also known as ``cosmic shear''.
The first conclusive detection of cosmic shear was
made in 2000 \cite{Wittman00}.

Fig.{\ref{fig:WL}} shows the constraints on
cosmological parameters $\sigma_8$ (the amplitude
of matter density fluctuations averaged on the 
scale of 8$\,h^{-1}$Mpc), $\Omega_m$, and a
constant dark energy equation of state $w=w_0$
from current weak lensing data \cite{Hoek06}.
Fig.{\ref{fig:WL_sig8}} shows how the measurement
of $\sigma_8$ has evolved over the last several years \cite{Hetter06}.
\begin{figure} 
\centerline{\epsfxsize=\figsize\epsffile{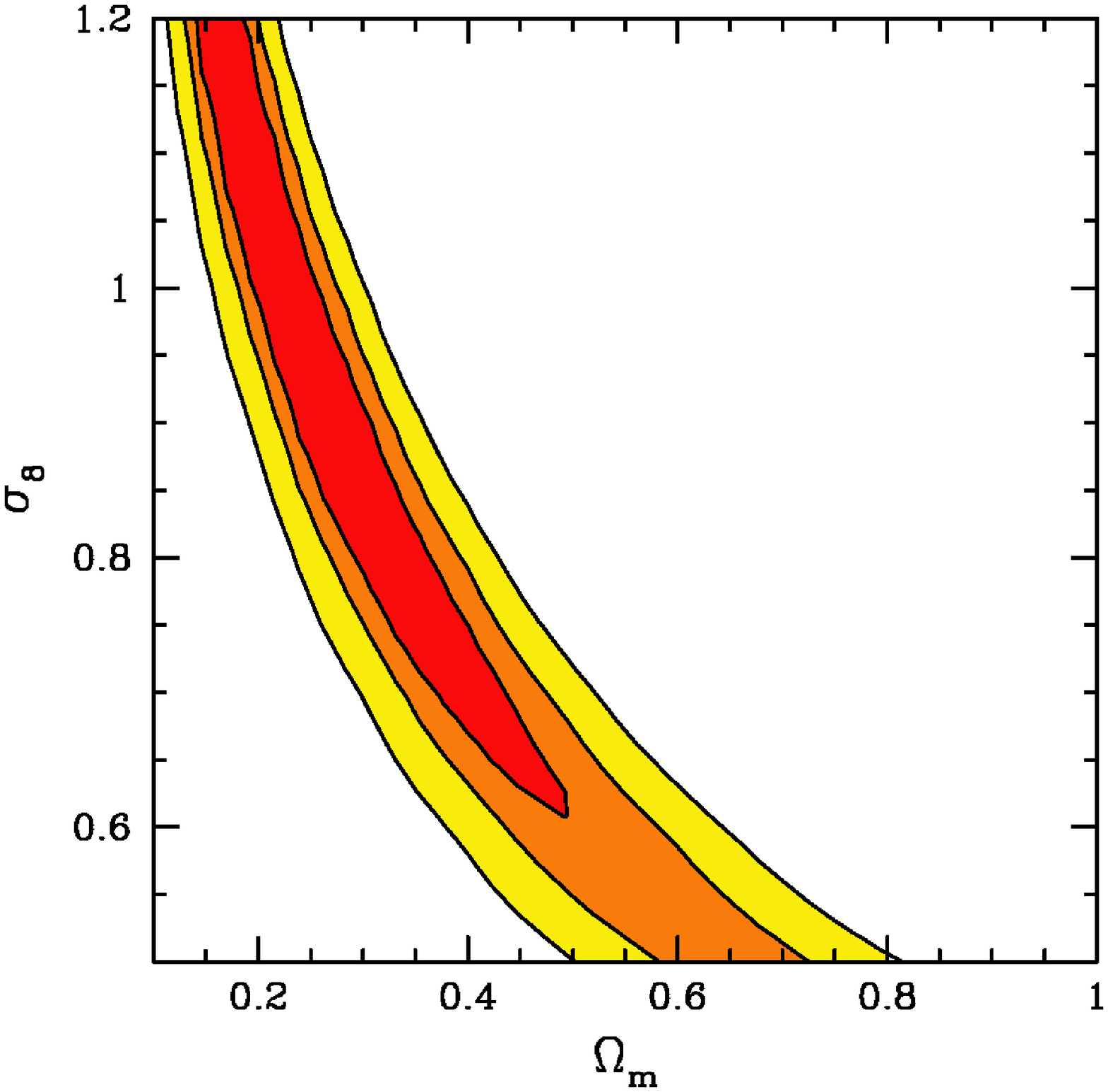}}
\centerline{\epsfxsize=\figsize\epsffile{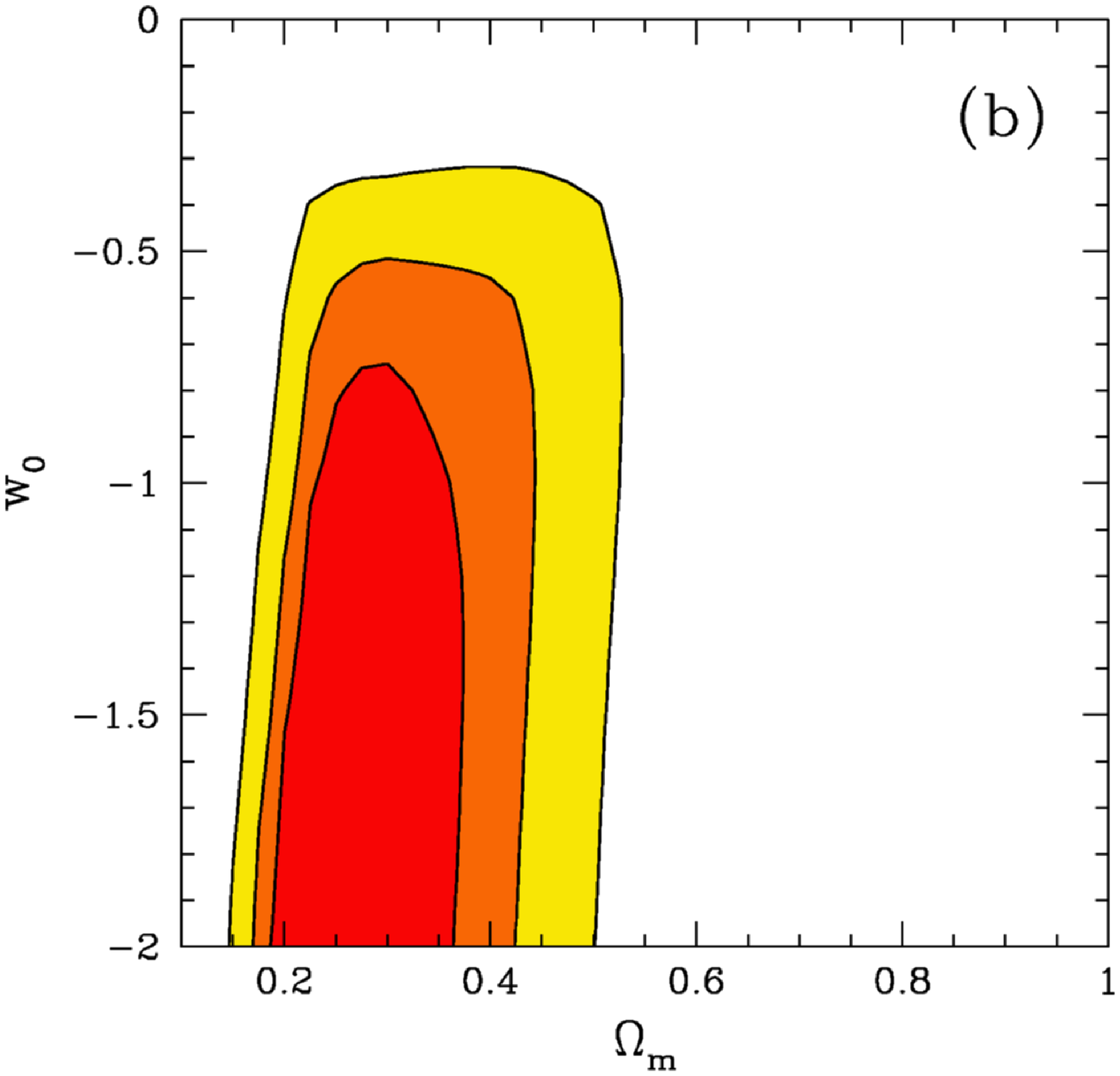}}
\vskip-0.3cm
\caption[1]{\label{fig:WL}\footnotesize%
Current constraints on cosmological parameters 
and dark energy from 
the CFHTLS Deep and Wide weak lensing survey,
assuming a constant $w_X(z)$ and a flat universe.
\cite{Hoek06}.
}
\end{figure}

\begin{figure} 
\centerline{\epsfxsize=\figsize\epsffile{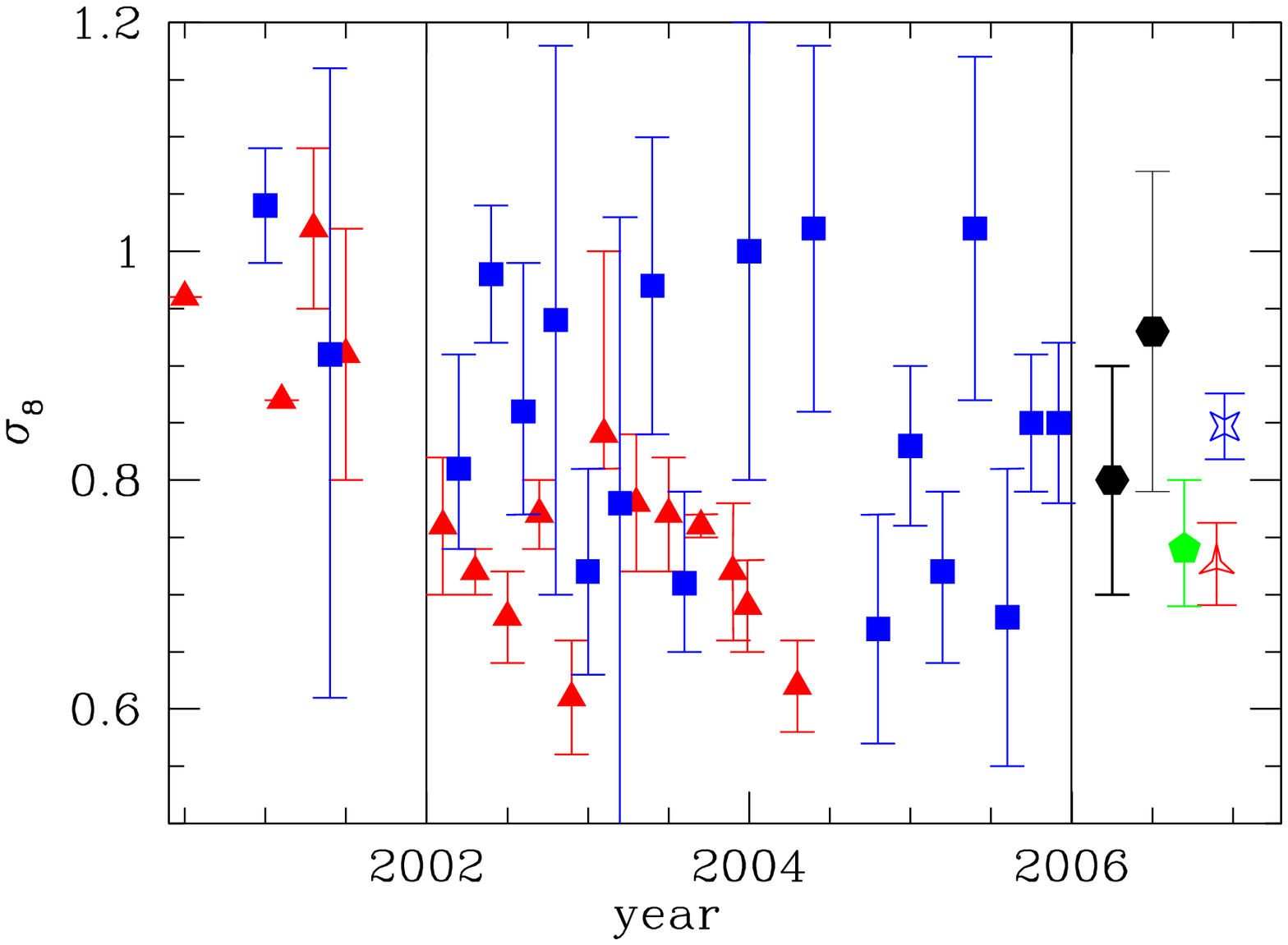}}
\caption[1]{\label{fig:WL_sig8}\footnotesize%
The measurements of $\sigma_8$ from 
the analysis of clusters of galaxies (red) and cosmological
weak lensing over the last several years.
\cite{Hetter06}
}
\end{figure}

Fig.{\ref{fig:WL}} and Fig.{\ref{fig:WL_sig8}} illustrate
both the advantage and the disadvantage of weak lensing
as dark energy probe.
weak lensing has the advantage of
being sensitive to the clustering of matter,
which can potentially allow us to differentiate
between dark energy and modified gravity \cite{Knox06}.
However, if the clustering of matter is not accurately
measured, the uncertainty is propagated into the
measurement of dark energy parameters (see Fig.{\ref{fig:WL}}).

If the systematic effects are properly modeled, a
weak lensing survey can potentially allow us to differentiate
between dark energy and modified gravity 
(see Fig.{\ref{fig:WL_g}}) \cite{Knox06}.
Fig.{\ref{fig:WL_g}} assumed a  
``2$\pi$'' deep and wide survey of
20,000 square degrees in six wavelength bands from $0.4-1.1$ $\mu$m 
to be undertaken by the Large Synoptic Survey Telescope (LSST).
This survey will yield  the shear and photometric redshift
of 3 billion source galaxies over a redshift range of 0.2 to 3.
\cite{Knox06}
\begin{figure} 
\centerline{\epsfxsize=\figsize\epsffile{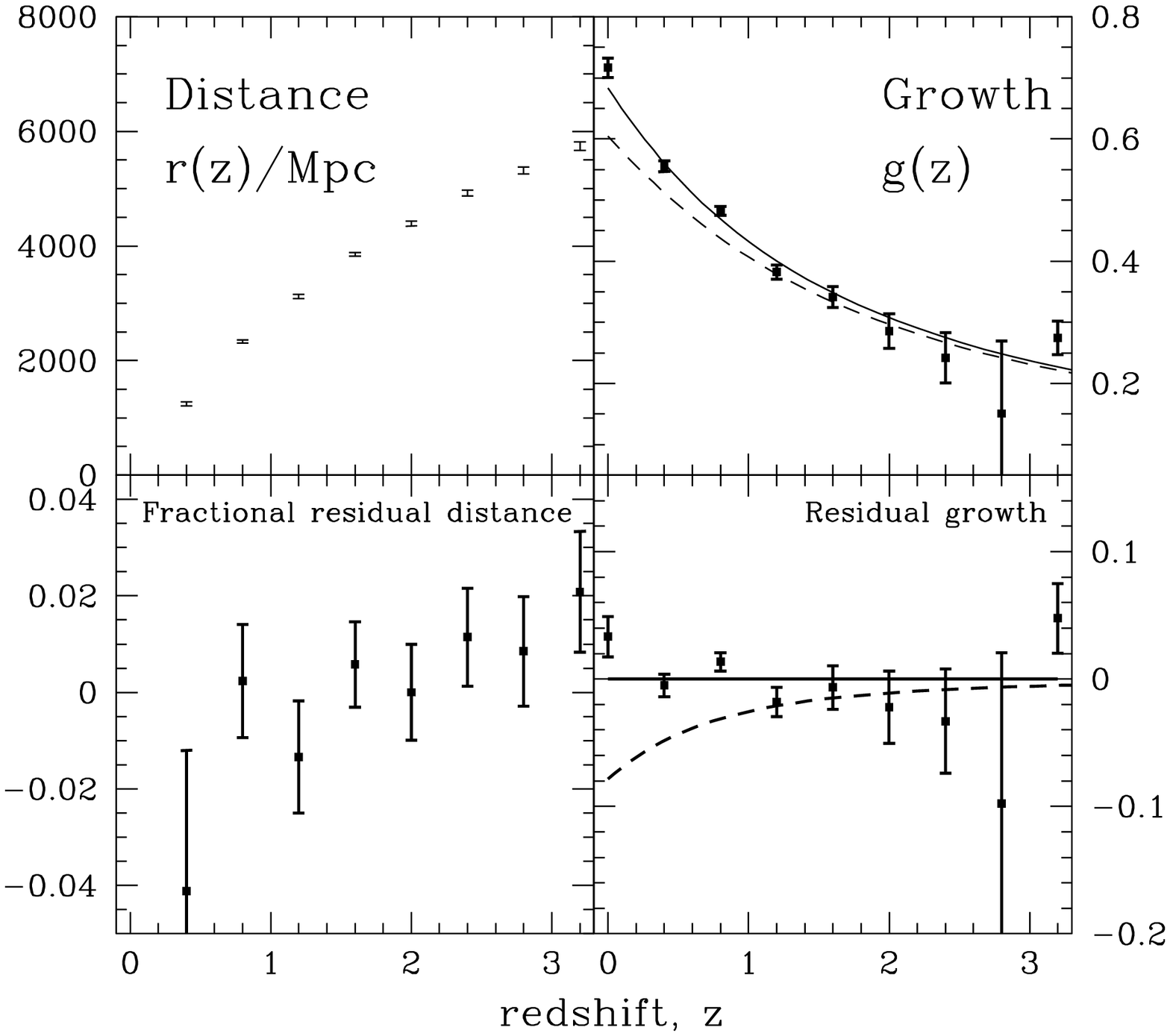}}
\caption[1]{\label{fig:WL_g}\footnotesize%
Reconstructed distances (left panels), 
and growth factors (right panels) from a LSST-like weak
lensing survey.  The lower left panel shows the 
fractional residual distances, 
$[r(z)-r_{\rm fid}(z)]/r_{\rm fid}(z)$, where $r(z)$ are the reconstructed
distances and $r_{\rm fid}(z)$ are the distances in the fiducial DGP model.
The lower right panel shows the residual growth factor, $g(z)-g_{\rm fid}(z)$.
The curves in the right panels are $g_{\rm fid}(z)$ (solid) and 
$g(z)$ for the Einstein gravity model 
(dashed) with the same $H(z)$ and $\rho_m$ as the DGP model.  Although
these two models have the same $r(z)$ they are distinguishable by
their significantly different growth factors. \cite{Knox06}.
Note that $g(z) \propto D_1(t)$ [see Eqs.(\ref{eq:fg})
and (\ref{eq:fg_DGP})].
}
\end{figure}

Weak lensing can be used to probe dark energy in two different
ways: weak lensing tomography \cite{Wittman00}, or
weak lensing cross-correlation cosmography \cite{Jain03}
(also known as ``the geometric method'' \cite{Zhang05}).
Much of the effort has been
focused on using the geometric weak lensing method to probe
dark energy to minimize the sensitivity to
the clustering of matter (which can be a source of
systematic uncertainty).

The basic idea of the geometric weak lensing method is to construct a map of the foreground
galaxies, from which an estimated map of the foreground mass can
be made. This foreground mass slice induces shear on all the galaxies
in the background. The amplitude of the induced shear as a function
of the background redshift is measured, from which 
a weighted sum of the ratios of
angular diameter distances between the source slice and lens slice, and 
between the lens slice and observer is estimated. \cite{Jain03,Zhang05}
Note that this marks an important difference of the weak lensing 
method from the supernova and baryon acoustic oscillation methods:
the weak lensing method gives correlated measurements
of the cosmic expansion history $H(z)$ in redshift bins,
while the supernova and baryon acoustic oscillation methods
can give uncorrelated measurements of $H(z)$ \cite{WangTegmark05,SE03}.

In the geometric weak lensing method, 
photometric redshifts\footnote{Photometric redshifts
are approximate redshifts estimated from multi-band imaging data.
They are calibrated using a spectroscopic sample.} are used
to divide galaxies into redshift bins.
The centroids of the photometric redshifts must be know to
the accuracy of around 0.1\% in order to avoid significant
degradation of dark energy constraints \cite{Bern04,Huterer06}
(see Fig.{\ref{fig:WL_degradation}}).
\begin{figure} 
\psfig{file=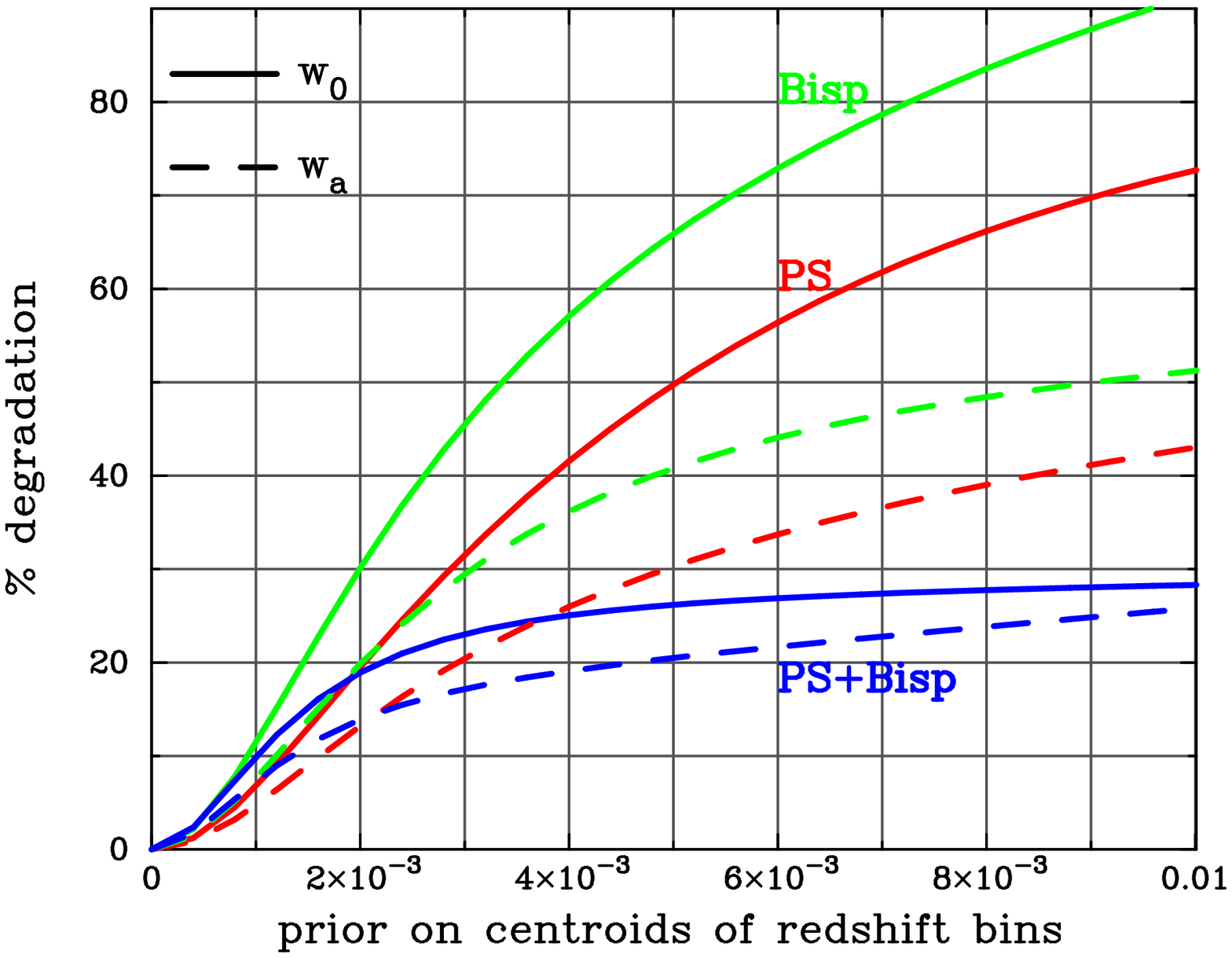,width=3.5in,angle=270} 
\caption[1]{\label{fig:WL_degradation}\footnotesize%
The centroids of the photometric redshifts must be known to
the accuracy of around 0.1\% in order to avoid significant
degradation of dark energy constraints \cite{Huterer06}.
}
\end{figure}

In addition to the uncertainty in the centroids of photometric redshift
bins, there are other systematic uncertainties in the use of
weak lensing as dark energy probe.
These include point spread function (PSF) correction,
bias in the selection of the galaxy sample, and
the intrinsic distortion signal due to the intrinsic alignment
of galaxies (see the detailed discussion in Ref.\cite{Peacock06}).

\subsection{Clusters as dark energy probe}

Clusters of galaxies can be used to probe dark energy in two different
ways: (1) using the cluster number density and its redshift evolution,
as well as cluster distribution on large scales (see
for example \cite{Haiman01,Vikh03,Schuecker03});
(2) using clusters as standard candles by assuming a constant cluster
baryon fraction (see for example, \cite{Allen04}), or using combined 
X-ray and SZ measurements for absolute distance measurements
(see for example \cite{Molnar04}).

Large, well-defined and statistically complete samples of galaxy 
clusters are required to derive rebust dark energy constraints
from cluster data. Future surveys aim to  
select clusters using data from X-ray satellite with high resolution 
and wide sky coverage, and multi-band optical and near-IR surveys to 
obtain photometric redshifts for clusters.\cite{Haiman01}

The systematic uncertainties of clusters as dark energy probe
include uncertainty in the cluster mass estimates that are 
derived from observed properties, such as X-ray or optical 
luminosities and temperature
(e.g. \cite{Majumdar03,Maju04,Lima04}).
Fig.{\ref{fig:cl}} shows the total cluster mass
versus a proxy based on the total baryon mass and temperature
(both of which can be inferred from X-ray observations),
based on simulated X-ray data \cite{Krav06}.
\begin{figure} 
\centerline{\epsfxsize=\figsize\epsffile{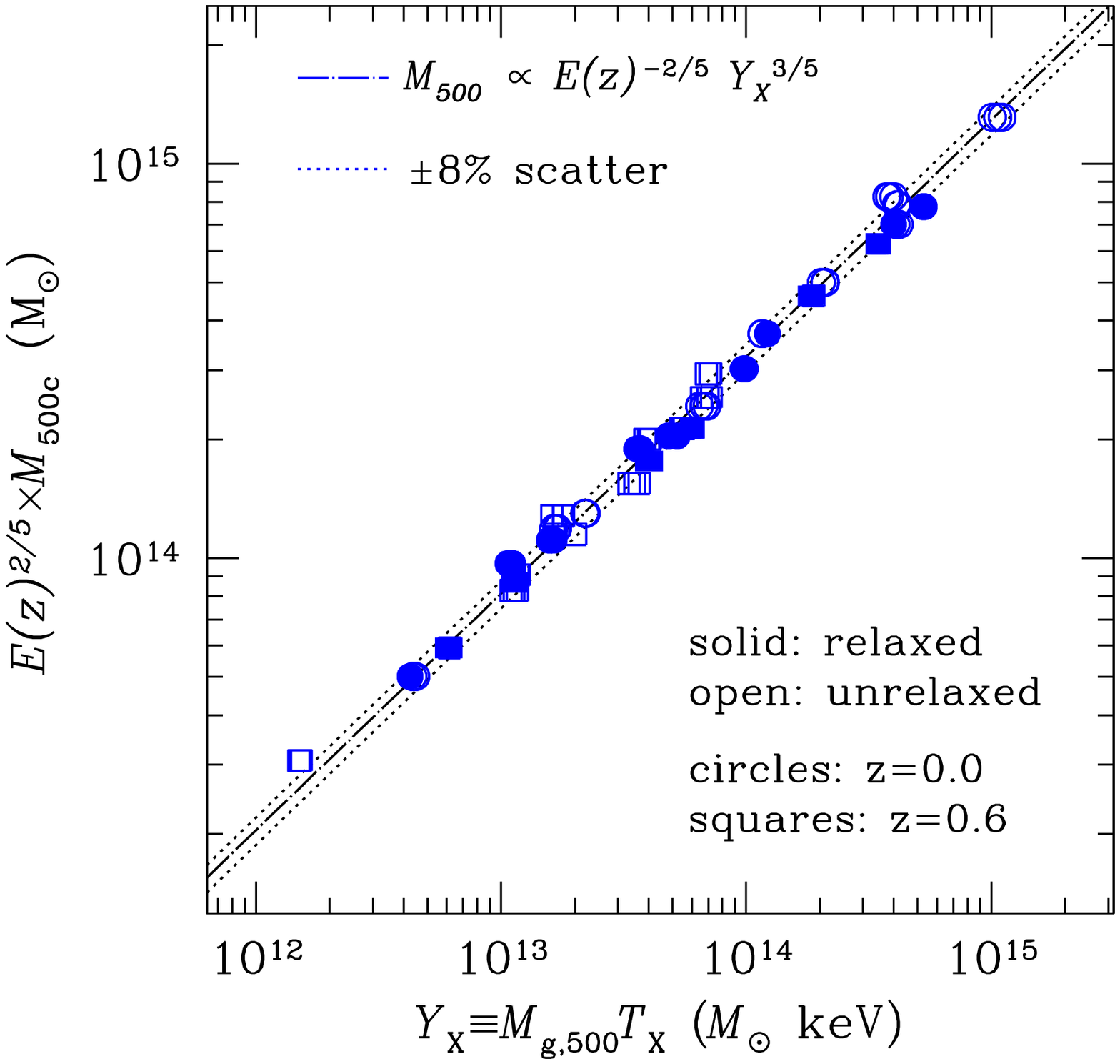}}
\caption[1]{\label{fig:cl}\footnotesize%
Total cluster mass versus a proxy based on the total 
baryon mass and temperature
(both of which can be inferred from X-ray observations),
based on simulated X-ray data \cite{Krav06}.
}
\end{figure}

\section{Current Status and Future prospects}

There are a large number of dark energy surveys that
are ongoing or have been proposed.
Ongoing projects include Essence \cite{Essence}, 
Supernova Legacy Survey (SNLS) \cite{SNLS},
Carnegie Supernova Project (CSP) \cite{CSP},
ESO Visible and Infrared
Survey Telescope for Astronomy (VISTA) Surveys \cite{VISTA}, 
Panoramic Survey Telescope \& Rapid Response System
(Pan-STARRS) \cite{Pan-STARRS}, 
and WiggleZ \cite{WiggleZ}.
Proposed near term projects include 
Advanced Liquid-mirror Probe for Astrophysics, Cosmology and Asteroids
(ALPACA) \cite{ALPACA};
Dark Energy Survey (DES) \cite{DES};
Hobby-Eberly Telescope Dark Energy Experiment (HETDEX) \cite{HETDEX};
Wide-Field Multi-Object Spectrograph \cite{wfmos}
and Sloan Digital Sky Survey (SDSS) III \cite{SDSSIII}.
Proposed long-term projects include 
Large Synoptic Survey Telescope (LSST) \cite{lsst},
Joint Dark Energy Mission (JDEM) \cite{jdem}, 
Square Kilometre Array (SKA) \cite{ska}, and an 
European-led dark energy mission \cite{esa}.

The U.S. Dark Energy Task Force (DETF) has recommended 
an aggressive, multi-stage, multi-method program to 
explore dark energy as fully as possible.\cite{Albrecht06}
DETF recommended a dark energy program with multiple techniques 
at every stage, with at least one of these being a probe sensitive to 
the growth of cosmic structure in the form of galaxies and 
clusters of galaxies.

The DETF defined a figure of merit that is
the inverse of the area of the 95\% confidence level
error ellipse in the $w_0$-$w_a$ plane [assuming
a dark energy equation of state $w_X(a)=w_0+w_a (1-a)$].
DETF recommended that dark energy program in Stage III 
(near-term, medium-cost projects) should be designed to achieve
at least a factor of 3 gain over Stage II (ongoing projects) 
in the figure of merit, and 
that dark energy program in Stage IV 
(long-term, high-cost projects JDEM, LST, SKA) should be
designed to achieve at least a factor of 10 gain over 
Stage II in the figure of merit.

The DETF recommended continued research and development 
investment to optimize JDEM, LST, and SKA (Stage IV) 
to address remaining technical questions and systematic-error risks, and 
high priority for near-term projects to improve understanding of dominant 
systematic effects in dark energy measurements, and wherever possible, reduce them.
The DETF recommended a coherent program of experiments designed to 
meet the goals and criteria it proposed.

The ESA-ESO Working Group on Fundamental Cosmology
made specific recommendations to ESA and ESO.
It recommended a wide-field optical and near-IR imaging survey
(suitable for weak lensing and cluster surveys) as a high priority, with 
ESA launching a satellite for high resolution wide-field optical 
and near-IR imaging, and ESO carrying out optical multi-color photometry,
as well as a large spectroscopic survey ($>$100,000 redshifts 
over $\sim$10,000 square degrees) to calibrate photometric redshifts.
They also recommended that ESA-ESO secure access to an instrument 
with capability for massive multiplexed deep spectroscopy (several thousand 
simultaneous spectra over one square degree) (suitable for
large galaxy redshift surveys), as well as conduct 
a supernova survey with multi-color imaging to extend existing 
samples of $z=0.5-1$ SNe by an order of magnitude, and improve 
the local sample of SNe. They suggested the 
use of a European Extremely Large Telescope (ELT) to study SNe at $z >1$. 

The U.S. National Research Council's Committee on
NASA's Beyond Einstein Program recently made the recommendation that
NASA and DOE should proceed immediately with a competition 
to select a JDEM for a 2009 new start. 
They concluded that ``The broad mission goals 
in the Request for Proposal should 
be (1) to determine the properties of dark energy with high 
precision and (2) to enable a broad range of astronomical 
investigations. The committee encourages the
Agencies to seek as wide a variety of mission 
concepts and partnerships as possible.''

ESA's Cosmic Vision 2015-2025 recently made the
first selection of candidate missions for assessment studies.
These include two mission concept for a European-led dark energy
mission: DUNE, the Dark UNiverse Explorer \cite{DUNE},
and SPACE, the SPectroscopic All-sky Cosmic Explorer \cite{SPACE}.
The synergy between JDEM and the ESA-led dark energy mission
will be important for a strategically optimized approach
to discovering the nature of dark energy.

In evaluating the various dark energy projects, it is critical to 
note that the challenge to solving the dark energy mystery will not be the 
statistics of the data obtained, but the tight control of 
systematic effects inherent in the data.
A combination of the most promising methods (as discussed in
this review), each optimized by having its systematics minimized by design,  
provides the tightest control of systematics \cite{JEDI}.
The discovery of the nature of dark energy
will revolutionize our understanding of the universe.

{\bf Acknowledgements:}
I am grateful to Daniel Eisenstein, Mario Hamuy, Marco Hetterscheidt,
Henk Hoekstra, Dragan Huterer, Lloyd Knox, Andrey Kravtsov,
and Kevin Krisciunas for permission to use their figures
in this review paper.

\end{document}